\documentclass[runningheads]{llncs}
\usepackage{graphicx}
\usepackage{amsmath}
\usepackage{bm}
\usepackage[utf8]{inputenc}
\usepackage{tikz}
\usetikzlibrary{positioning}
\usetikzlibrary{calc}
\usetikzlibrary{arrows}
\usetikzlibrary{arrows.meta}
\usepackage{booktabs}
\usepackage{url}

\usepackage{color, colortbl}

\begin{document}
\title{Non-Rigid Volume to Surface Registration using a Data-Driven Biomechanical Model}
\titlerunning{Non-Rigid Volume to Surface Registration}
%
\author{Micha Pfeiffer\inst{1} \and
Carina Riediger\inst{2} \and
Stefan Leger\inst{1} \and
Jens-Peter Kühn\inst{3} \and
Danilo Seppelt\inst{3} \and
Ralf-Thorsten Hoffmann\inst{3} \and
Jürgen Weitz\inst{2} \and
Stefanie Speidel\inst{1}}
\authorrunning{M. Pfeiffer et al.}
%
\institute{Translational Surgical Oncology, National Center for Tumor Diseases, Dresden, Germany (\email{micha.pfeiffer@nct-dresden.de}) \and
Department for Visceral, Thoracic and Vascular Surgery, University Hospital Carl-Gustav-Carus, TU Dresden, Germany \and
Institute and Policlinic for Diagnostic and Interventional Radiology, University Hospital Carl-Gustav-Carus, TU Dresden, Germany}
\maketitle              
\begin{abstract}
Non-rigid registration is a key component in soft-tissue navigation. We focus on laparoscopic liver surgery, where we register the organ model obtained from a preoperative CT scan to the intraoperative partial organ surface, reconstructed from the laparoscopic video.
This is a challenging task due to sparse and noisy intraoperative data, real-time requirements and many unknowns - such as tissue properties and boundary conditions.
Furthermore, establishing correspondences between pre- and intraoperative data can be extremely difficult since the liver usually lacks distinct surface features and the used imaging modalities suffer from very different types of noise.
In this work, we train a convolutional neural network to perform both the search for surface correspondences as well as the non-rigid registration in one step. The network is trained on physically accurate biomechanical simulations of randomly generated, deforming organ-like structures. This enables the network to immediately generalize to a new patient organ without the need to re-train. We add various amounts of noise to the intraoperative surfaces during training, making the network robust to noisy intraoperative data. During inference, the network outputs the displacement field which matches the preoperative volume to the partial intraoperative surface. In multiple experiments, we show that the network translates well to real data while maintaining a high inference speed. Our code is made available online.


\end{abstract}

\section{Introduction}

\begin{figure}[ht]
\begin{tikzpicture}
\centering
\node[above right] (img) at (0,0) {
\includegraphics[width=0.99\textwidth]{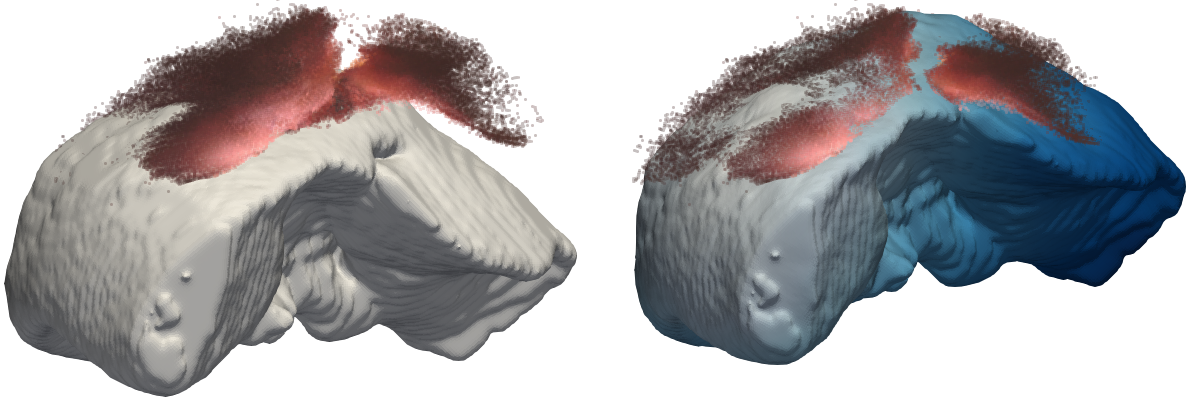}};
\node[anchor=south west] at ($(img.south west)+(0,0)$) {a)};
\node[anchor=south west] at ($(img.south west)+(0.5\textwidth,0)$) {b)};
\end{tikzpicture}
\caption{Non-rigid registration using our CNN. Given the intraoperative point cloud of a partial liver surface and the preoperative liver mesh (a), our network estimates a displacement field, which deforms the preoperative volume to match the intraoperative surface (b). Displacement magnitude is encoded in shades of blue.}
\label{fig:LaparoscopicResult}
\end{figure}

In navigated surgical interventions, the aim is to aid surgeons in finding structures of interest - such as tumors, vessels or planned resection lines.
Often, a detailed, accurate and highly informative preoperative computer tomography (CT) scan is available and the challenge is to align this model with the intraoperative scene. Whenever deforming soft-tissue is involved, this requires a non-rigid registration.
However, usually, only limited, sparse and noisy data can be acquired during the intervention.
Furthermore, it is usually very difficult to find one-to-one correspondences between the pre- and intraoperative data, since they are subject to different types of noise and can look substantially different.
Together with the large deformation, only partially visible surface and many unknown parameters - such as the organ's elastic properties, acting forces and boundary conditions - this makes the soft-tissue registration a challenging problem.

This work explores a deep-learning based approach to performing a non-rigid organ registration. We focus on laparoscopic liver surgery and register a given preoperative liver volume mesh to an intraoperative, partial surface of the same organ (obtained from a laparoscopic stereo video stream). We use a fully convolutional neural network (CNN), which analyses the two meshes and outputs a displacement field to register the preoperative organ volume to the intraoperative surface (see Fig. \ref{fig:LaparoscopicResult}). To reach this aim, the network must learn to find a plausible solution for the surface correspondence problem, while at the same time learning about biomechanical deformation in order to constrain itself to physically realistic solutions.
For training, a deformed intraoperative state of each preoperative mesh is simulated, using the Finite Element Method (FEM). We use synthetically generated, random organ-like meshes, which allows the trained network to generalize to new patients without the need to re-train for every liver.
We call the approach \emph{Volume-To-Surface Registration Network} (V2S-Net) and publish the code as well as the pretrained network online\footnote{\url{https://gitlab.com/nct_tso_public/Volume2SurfaceCNN}}.


\subsubsection{Related Work}
The application of deep learning to simulate organ deformation has recently received a lot of attention, mainly due to the very low inference times \cite{Mendizabal2019,MendizabalTagliabue2019,PellicerValero2020,Pfeiffer2019}. It has been shown that these data-driven models can achieve a high accuracy \cite{Mendizabal2019,PellicerValero2020}, can deal with partial surface information \cite{Brunet2019,Pfeiffer2019} and can even learn to deform real organs after training on synthetic data only \cite{Pfeiffer2019}. 
However, all of the mentioned methods require the use of boundary conditions similar to the FEM. These are very difficult to obtain in a real surgical setting, since, for example, forces are extremely difficult to measure and estimating the surface displacement would require known surface correspondences.

On the other hand, the machine learning community has developed a large number of data-driven feature descriptors to efficiently describe and register 3D structures \cite{DeepLearning3DReview2019,3DKeypointDescriptors2018,PointNet2017}. While this shows the ability of neural networks to interpret 3D data, in the case of organs, the lack of distinct features and large organ deformation requires the incorporation of soft-tissue mechanics in order to find correct correspondences. 

In design, our method is similar to the work of Suwelack et al. \cite{Suwelack2014}, who propose to morph the preoperative volume mesh into the intraoperative partial surface by simulating electric charge on the surfaces and solving the problem using the FEM. Similar to our work, they use the two meshes as input and output a displacement field for the preoperative mesh. However, like many previous approaches, their method requires manual assignment of boundary conditions and parameters, can become unstable if wrong values are chosen and their displacement estimation is much slower than ours.



\section{Methods}
The aim of the proposed V2S-Net is to learn to recover a displacement field which deforms the
input volume in such a way that it is well aligned with the partial surface, when given only the preoperative volume mesh and a partial intraoperative surface as input.
We train the network on synthetic data, since a real data set consisting of volume meshes as well as known displacement fields is not available. 
In addition to simulating deformations, we also generate the preoperative 3D meshes randomly, to ensure that the network will translate directly to new, unseen liver meshes.

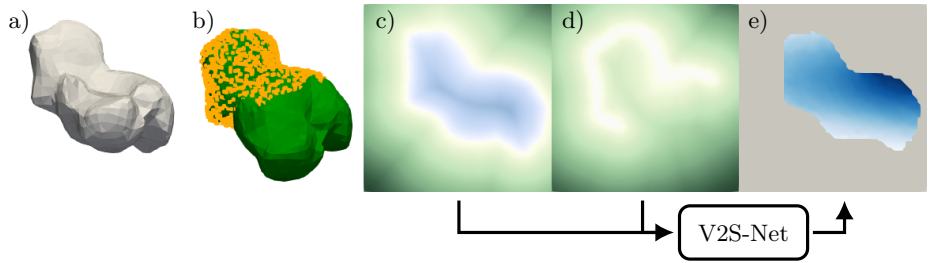
\begin{figure}[ht]
\begin{tikzpicture}
\node[above right] (img) at (0,0) {\includegraphics[width=\textwidth]{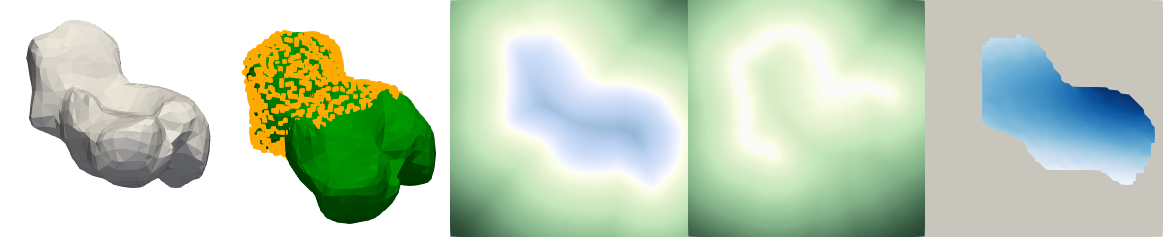}};
\node[anchor=north west] at ($(img.north west)+(0,-2pt)$) {a)};
\node[anchor=north west] at ($(img.north west)+(0.2\textwidth,-2pt)$) {b)};
\node[anchor=north west] at ($(img.north west)+(0.4\textwidth,-2pt)$) {c)};
\node[anchor=north west] at ($(img.north west)+(0.6\textwidth+1pt,-2pt)$) {d)};
\node[anchor=north west] at ($(img.north west)+(0.8\textwidth+2pt,-2pt)$) {e)};
\node[anchor=north,very thick,draw,rounded corners,inner sep=7,outer sep=2] (CNN) at (0.81\textwidth,0) {V2S-Net};
\draw[very thick,-{Latex[length=2.5mm,width=2.5mm]}] (0.5\textwidth,0) |- (CNN.west);
\draw[very thick] (0.7\textwidth,0) |- ($(CNN.west)-(1mm,0)$);
\draw[very thick,-{Latex[length=2.5mm,width=2.5mm]}] (CNN.east) -| (0.92\textwidth,0);

\end{tikzpicture}
\caption{a) Random preoperative volume mesh $\text{V}_P$, b) simulated intraoperative state $\text{V}_I$ (green) and partial surface $S_I$ (orange), c) signed distance field of preoperative volume $\text{SDF}_P$, d) distance field of partial surface $\text{DF}_I$, e) known displacement $\text{U}$ (magnitude), which the V2S-Net learns to predict from the two distance fields. Afterwards, this displacement field can be used to infer the position of internal structures, such as vessels and tumors.}
\label{fig:DataGeneration}
\end{figure}

\subsection{Data Generation}
\label{sec:DataGeneration}

\subsubsection{Simulation}

A random 3D surface mesh is generated by first creating an icosphere and then applying a set of extrusion, boolean, remeshing and smoothing operators. We use \textit{Gmsh} \cite{Gmsh2009} to fill the surface mesh with tetrahedral elements. The resulting volume mesh is considered to be our preoperative volume organ mesh $\text{V}_{P}$ (Fig. \ref{fig:DataGeneration}, a). Next, up to three random forces (max. magnitude 1.5 N) and a zero-displacement boundary condition are assigned to random areas of the surface of the mesh. The steady-state deformation is calculated by the \textit{Elmer} \cite{Elmer2013} finite element solver, using a neo-Hookean hyperelastic material model with a random Young's Modulus (2 kPa to 5 kPa) and Poisson's ratio of 0.35. The resulting deformed volume acts as the intraoperative state of the organ $\text{V}_{I}$ (Fig. \ref{fig:DataGeneration} b). For every vertex in $\text{V}_{P}$, we now know the displacement vector that needs to be applied in order to reach the intraoperative state $\text{V}_{I}$, resulting in the displacement field $u$.


We extract a random surface patch of this deformed volume mesh. To simulate intraoperative noise and difference in imaging modalities, we resample this patch, delete random vertices and add uniform random displacements to the position of every vertex. Furthermore, random parts of the patch are deleted to simulate areas where surrounding tissues (like the falciform ligament) occlude the surface from the perspective of the laparoscope. The result is our intraoperative partial surface mesh $\text{S}_{I}$.

The use of random values sometimes leads to the generation of meshes for which the creation of tetrahedral elements fails or simulations for which the finite element solver does not find a solution. These samples are discarded. We also discard samples where the maximum displacement is larger than 20 cm or the amount of visible surface is less than 10\%, since we assume these cases to be unrealistic.

\subsubsection{Voxelization}

To pass the meshes to the network, we represent them in the form of distance fields on a regular grid. For this, we define a grid of $64^3$ points. At each point, we calculate the distance to the nearest surface point of the preoperative volume $\text{V}_P$ as well as the intraoperative surface $\text{S}_I$. For the preoperative mesh, we flip the sign of all grid points that lie within the organ volume, resulting in a signed distance field $\text{SDF}_P$ for the preoperative and a distance field $\text{DF}_I$ for the intraoperative mesh (see Fig. \ref{fig:DataGeneration} c and d).

Additionally, we interpolate the target displacement field $u$ into the same grid with a gaussian kernel, resulting in an interpolated vector field $\text{U}$. For points outside the preoperative volume, $\text{U}$ is set to $(0,0,0)^T$.

We use the outlined process to generate $80~000$ samples, which are split into training data (90\%) and validation data (10\%). By flipping each training sample along each combination of X-, Y- and Z-axes, we further increase the amount of training data by a factor of eight, resulting in roughly $460~000$ samples.

\subsection{Network Architecture and Training}

To estimate the displacement field $\text{U}$, the network is given the full preoperative volume $\text{SDF}_P$ and the partial intraoperative surface $\text{DF}_{I}$. Formally, we want our network to estimate the function:
\begin{equation}
F(\text{SDF}_{P},\text{DF}_{I}) = \text{U}.
\end{equation}

Similar to \cite{Mendizabal2019} and \cite{Pfeiffer2019}, we choose a fully convolutional architecture with 3D convolutions, an encoder-decoder structure and skip connections to allow for an abstract low-resolution representation while preserving high-resolution details. The precise architecture is shown in the supplementary material. Following \cite{Pfeiffer2019}, we let the network output the displacement field $\text{U}_{r,est}$ at multiple resolutions $r \in (64, 32, 16, 8)$. These are compared to the (downsampled) target displacement fields $\text{U}_r$ using the mean absolute error $L_r$. We find that this process speeds up loss convergence during training. The final loss $L$ is a weighted sum of these errors:
\begin{equation}
\begin{split}
L_r &= \frac{1}{r^3} \sum_{i=1}^{r^3} { \left| \text{U}_r(i) - \text{U}_{r,est}(i) \right| } \\
L &= 10~ L_{64} + L_{32} + L_{16} + L_{8}.
\end{split}
\end{equation}

We train the network with a \textit{one cycle learning rate scheduler} \cite{superconvergence2017} and the \textit{AdamW} optimizer \cite{adamW2017} for 100 epochs, after which the mean registration error on the validation data has converged to roughly 6 mm.

\section{Experiments and Results}

Since our network is trained on randomly generated meshes, it is vital to test on real patient data. However, reference data for real laparoscopic interventions is very difficult to obtain, since interventional CT scans during laparoscopy are very limited. To nevertheless capture all necessary aspects of the registration process, we conduct three experiments: one experiment with simulated liver deformations, one with real human liver deformations under breathing motion and one qualitative experiment on data from a laparoscopic setting.
In all experiments, the used preoperative liver meshes were extracted from patient CT scans automatically using a segmentation CNN \cite{LiTS}.
The process of estimating the displacement field (including upload to the GPU) takes roughly 130 ms.

\subsection{Liver Deformations (In Silico)}

\begin{figure}[ht]
\includegraphics[width=\textwidth]{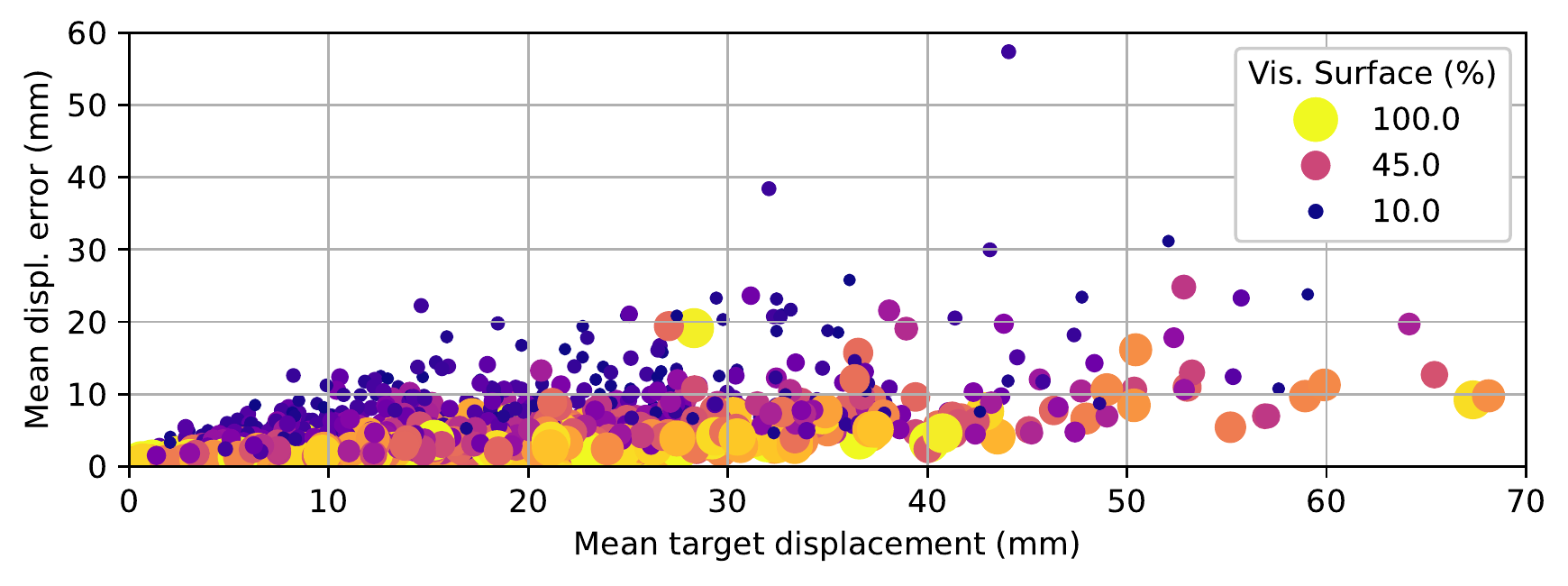}
\caption{Mean displacement error of each sample over the mean target displacement. As expected, the registration error tends to increase with target displacement and decrease with larger visible surfaces. With few exceptions, samples with a displacement error larger than 1 cm are those with very little visible surface.}
\label{fig:SimulatedDataStats}
\end{figure}

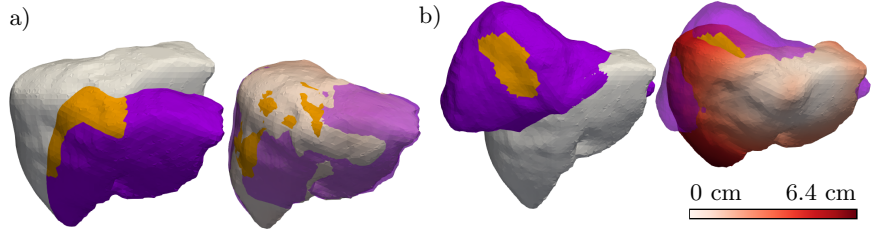
\begin{figure}[ht]
\centering
\begin{tikzpicture}
\node[left] (success) at (0,0) {\includegraphics[width=0.45\textwidth]{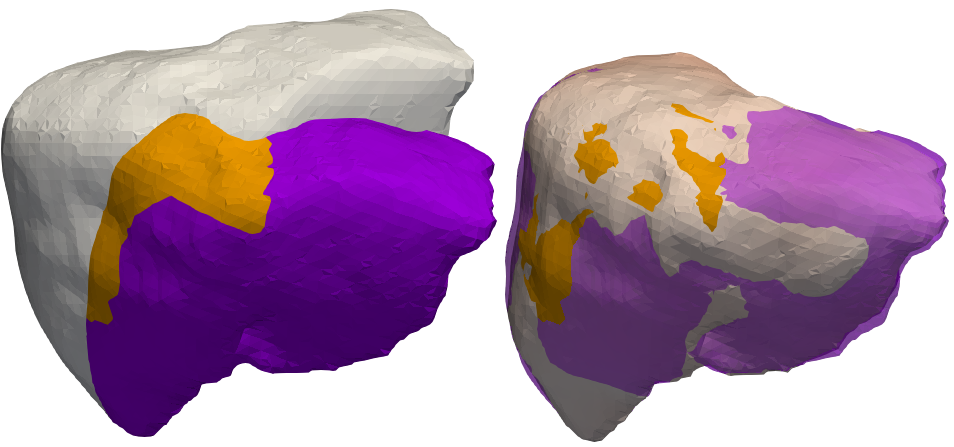}};
\node[anchor=north west] (failure) at ($(success.north east)+(0,0.5cm)$)
 {\includegraphics[width=0.47\textwidth]{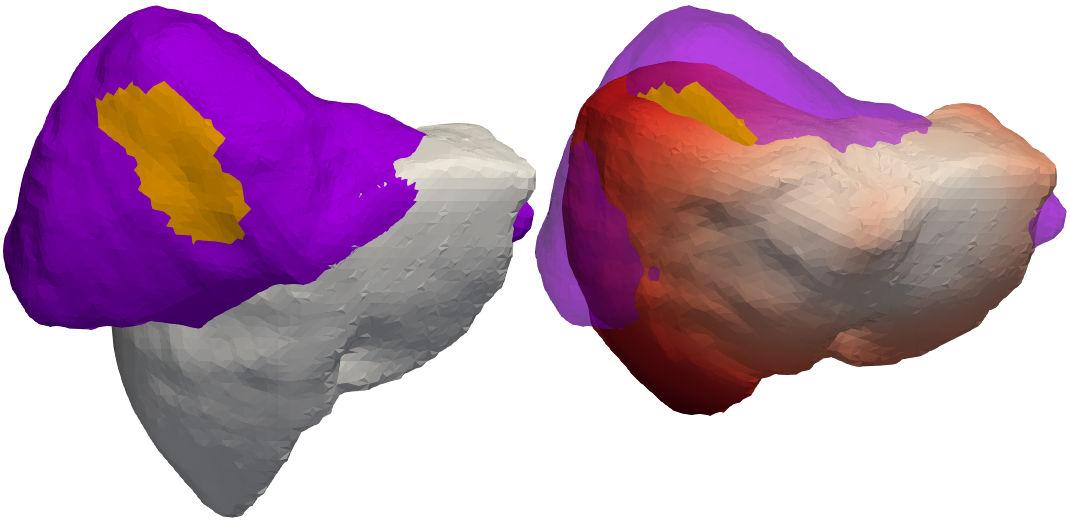}};

\node[anchor=south east,draw=black,inner sep=0,very thick, xshift=-3mm] (colorbar) at (failure.south east) {\includegraphics[width=0.18\textwidth,height=1mm]{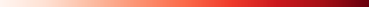}};
\node[anchor=south west,xshift=-1mm] at (colorbar.north west) {0 cm};
\node[anchor=south east,xshift=1mm] at (colorbar.north east) {6.4 cm};

\node[anchor=north west] at ($(success.north west)+(0,+0.4)$) {a)};
\node[anchor=north west] at ($(failure.north west)+(-8pt,-2pt)$) {b)};

\end{tikzpicture}
\caption{
From a real preoperative liver mesh (grey), a deformed intraoperative liver state (purple) is simulated. Given the preoperative mesh and a partial intraoperative surface (yellow), the V2S-Net estimates the deformed state (images 2 and 4).
The magnitude of the error (red) indicates a successful registration for case a) (max. error 1.5 cm) and an unsuccessful registration for case b) (max. error 6.4 cm).
When the randomly picked visible surface contains no distinct features (as is the case in b), or captures no areas of significant deformation, the network may estimate an incorrect but plausible deformation, resulting in the outliers in Fig. \ref{fig:SimulatedDataStats}. This suggests that it is more important which areas of the liver are visible than the size of the visible surface.}
\label{fig:SimulatedDataSamples}
\end{figure}

We generate a synthetic dataset as described in section \ref{sec:DataGeneration}. However, instead of generating random meshes we use the mesh of a patient's liver and simulate 1725 deformations and partial intraoperative surfaces. This new dataset is not used for training and thus allows us to test how our method translates to a real liver mesh: We apply the network to all samples and calculate the displacement error for every point in the estimated displacement fields $\text{U}_{64,est}$ (results in Fig. \ref{fig:SimulatedDataStats}). When the visible surfaces are large enough, registration errors tend to be small, even when the target deformations become large. Smaller visible areas lead to more interesting results, for example when the network cannot infer which part of the volume should be matched to the partial surface (see Fig. \ref{fig:SimulatedDataSamples}).

\subsection{Liver with Breathing Deformation (In Vivo)}





\begin{figure}[ht]
\centering
\includegraphics[width=0.9\textwidth]{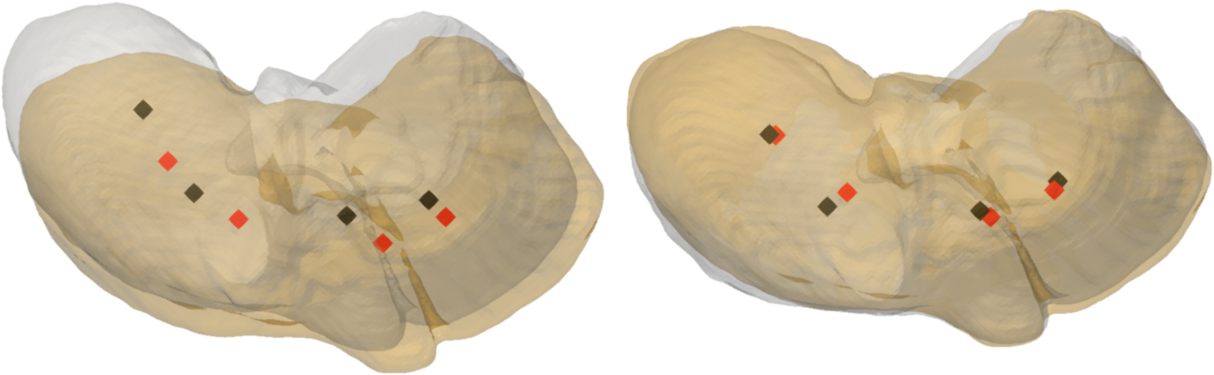}
\caption{Registration for Patient 1. The liver is deformed by breathing motion (inhaled and exhaled states, left). Four vessel bifurcations are marked in the first CT scan (black) and in the second CT scan (red). The network uses the surfaces to calculate a displacement field which registers the two liver states to each other (right). Applying this displacement field to the markers indicates a good approximation of the internal deformation. Results for Patient 2 can be found in the supplementary material.}
\label{fig:HumanBreathingMotionDD}
\end{figure}

During breathing motion, the human liver moves and deforms considerably.
To assess whether our network translates to real human liver deformation, we evaluate on human breathing data.
We extract liver meshes from two CT scans (one showing an inhaled and one an exhaled state) and let these represent the preoperative volume $V_P$ and intraoperative surface $S_I$. We search for clearly visible vessel bifurcations and mark their positions in both scans.
Given the distance fields of the two meshes, the network then estimates the displacement of every voxel in the inhaled state. The resulting displacement field $\text{U}_{64,est}$ is interpolated to the positions of the marked bifurcation points (gaussian interpolation kernel, radius 1 cm) and is used to displace them.
We carry out this experiment for two patients. Average displacements and remaining errors after registration are shown in Table \ref{tab:BreathingExperiment} and the result for Patient 1 is depicted in Fig. \ref{fig:HumanBreathingMotionDD}.

\begin{table}
\centering
\caption{Displacement of marked bifurcations (due to inhaling) and remaining errors after registration.}
\begin{tabular}{l@{\hspace{2em}}l@{\hspace{2em}}l}
& Avg. displacement (mm) & Avg. error (mm) \\  \midrule
Patient 1 & 21.4 (max: 31.1) & 5.7 (max: 9.8) \\
Patient 2 & 28.3 (max: 32.7) & 4.8 (max: 5.4) \\ \bottomrule
\end{tabular}
\label{tab:BreathingExperiment}
\end{table}

%



\subsection{Laparoscopic Liver Registration (In Vivo)}

To validate whether our method works for our target task of navigated laparoscopy, we perform a qualitative evaluation on human in-vivo data acquired from a da Vinci (Intuitive Surgical) stereo laparoscope. We first reconstruct the intraoperative surface $S_I$ from the video stream. For this, a 10 second camera-sweep of the liver is performed at the beginning of the surgery. Since we lack positional information of the laparoscope, the OrbSLAM2 \cite{ORBSLam2} algorithm is used to estimate the camera pose for each frame. Furthermore, the disparity between each pair of stereo frames is estimated (using a disparity CNN \cite{disparityCNN2019}) and a semantic segmentation is performed on each left camera frame to identify pixels which show the liver (using a segmentation CNN \cite{Pfeiffer2019Generating}). We reproject the pixel information into the scene using the disparity, camera calibration parameters and estimated camera pose, obtaining a 3D point cloud with color and segmentation information. While the camera moves, points which are close together and are similar in hue are merged together. After the sweep, we discard points if they have not been merged multiple times (i.e. points which were not seen from multiple view points) or if they were not segmented as \emph{liver} in at least 70\% of the frames in which they were seen. Additionally, we use a moving least squares filter (radius 0.5 cm) to smoothen the surface.
After a manual rigid alignment of $V_P$ and $S_I$, the distance fields for the pre- and intraoperative data are calculated and our network is used to estimate a displacement of the liver volume. Qualitative results are shown in Fig. \ref{fig:LaparoscopicResult} and in the supplementary material.

\section{Conclusion}

In this work, we have shown that a CNN can learn to register a full liver surface to a deformed partial surface of the same liver. The network was never trained on a real organ or real deformation during the training process, and yet it learned to solve the surface correspondence problem as well as the underlying biomechanical constraints.

Since our method works directly on the surfaces, no assignment of boundary conditions and no search for correspondences is necessary. The breathing motion experiment shows that, even though the segmentation process creates some artifacts in the surfaces, the V2S-Net finds a valid solution. The network was able to find the displacement without the need for a prior rigid alignment, likely because the full intraoperative surface was used. In cases with less visible surface, we find that the solution depends on the prior rigid alignment, which should be investigated further in future work.

In our method, we outsource the complex and time consuming simulation of soft-tissue behavior to the data generation stage. This could be further exploited by adding additional information to the simulations, such as inhomogeneous material properties, surrounding and connecting tissues and more complex boundary conditions. Where these properties are measurable for a patient, they could be passed to the network as additional input channels.

Our results show that there may be cases where the solution is ambiguous, for example when too little information is given and the network must guess how the hidden side of the liver is deformed.
This issue is likely inherent to the ill-posed laparoscopic registration problem in itself and not confined to our method. However, in contrast to conventional registration methods, neural networks can estimate how confident they are of a solution \cite{BayesianDL2016}. This probabilistic output could be used to assess how a solution should be interpreted and could give additional information to the surgeon.

%
%
\bibliographystyle{splncs04}
\bibliography{literature.bib}
\end{document}


\title{Non-Rigid Volume to Surface Registration using a Data-Driven Biomechanical Model: Supplementary Material}
\titlerunning{Non-Rigid Volume to Surface Registration}
%
\author{Micha Pfeiffer\inst{1} \and
Carina Riediger\inst{2} \and
Stefan Leger\inst{1} \and
Jens-Peter Kühn\inst{3} \and
Danilo Seppelt\inst{3} \and
Ralf-Thorsten Hoffmann\inst{3} \and
Jürgen Weitz\inst{2} \and
Stefanie Speidel\inst{1}}
\authorrunning{M. Pfeiffer et al.}
%
\maketitle              

\begin{figure}
\centering
\begin{tikzpicture}
\node(topview) at (0,0) {\includegraphics[width=0.8\textwidth]{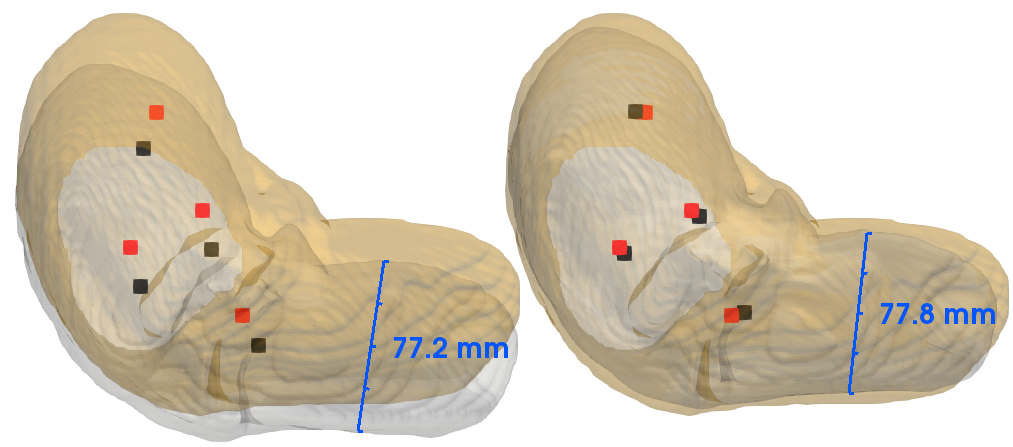}};
\node[anchor=north](frontview) at (topview.south) {\includegraphics[width=0.8\textwidth]{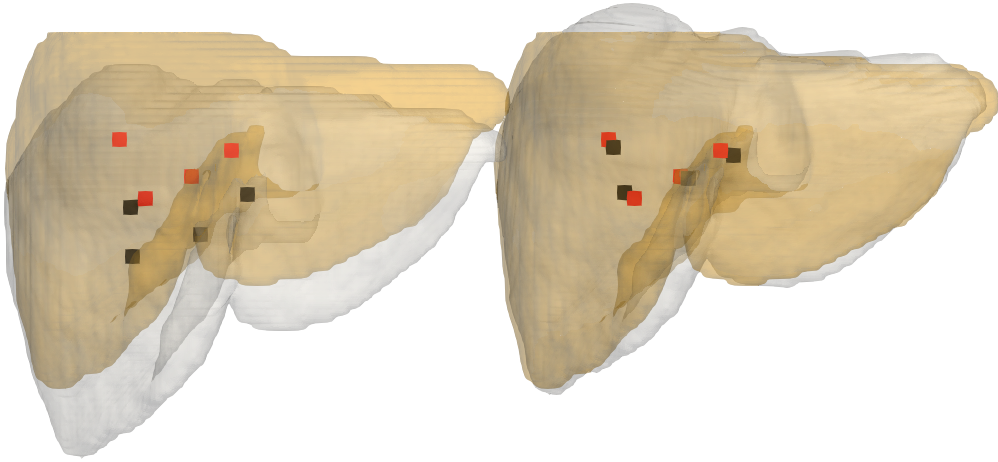}};
\draw[draw=black,->,thick] ($(frontview.north west)+(1.5cm,0)$) -- ($(frontview.north west)+(2cm,-0.7cm)$){};
\end{tikzpicture}
\caption{Breathing motion experiment, Patient 2. Left: Surfaces and positions of (corresponding) bifurcations extracted from CT. Right: Registration result.
Top row shows a transverse view, bottom row a coronal view.
One of the CT scans was cut off, thus one liver surface is missing a small part of the right  lobe (black arrow). Since the network was trained on varying amounts of visible surface, this poses no problem for our method, and a valid registration result is obtained. The segmentation process results in slightly smaller structures for the exhaled state of the liver, still the network manages to preserve them correctly during registration (blue rulers).}
\end{figure}

\begin{figure}
\begin{tikzpicture}
\tikzstyle{lbl} = [rectangle, draw=black, fill=white, inner sep=1mm, thin]
\tikzstyle{arr} = [->,draw=green,very thick]

\node(intraop_full) at (0,0) {\includegraphics[width=0.5\textwidth]{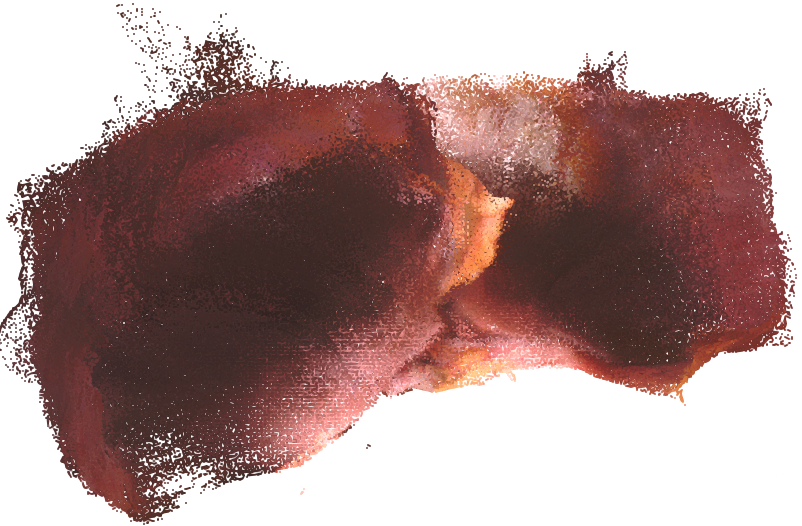}};

\node[anchor=north west](intraop_mc) at (intraop_full.north east) {\includegraphics[width=0.33\textwidth]{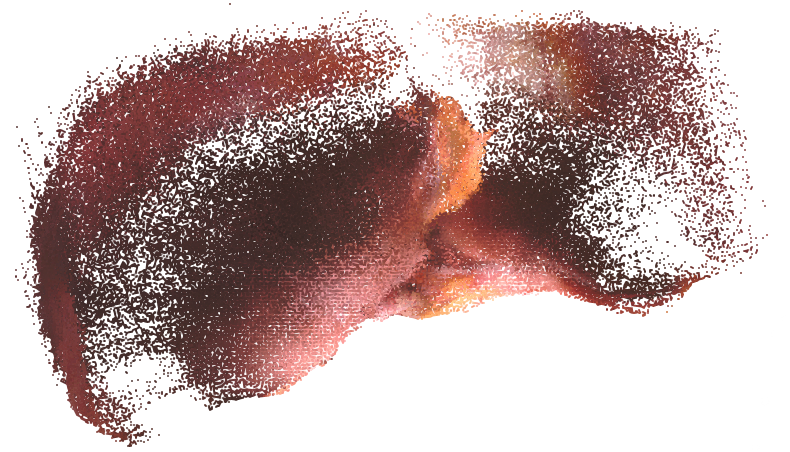}};
\node[anchor=south west](intraop_mc_seg) at (intraop_full.south east) {\includegraphics[width=0.33\textwidth]{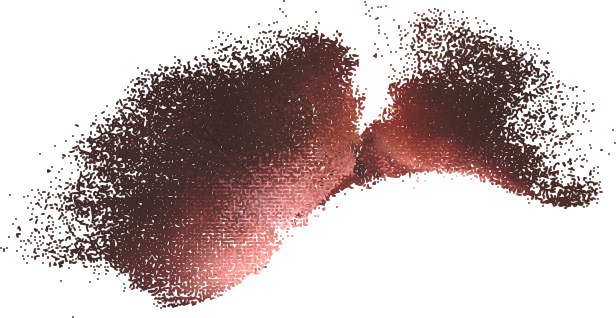}};

\node[lbl, anchor=north](ligament) at ($(intraop_full.north)+ (1cm,0cm)$)  {Falciform Ligament};
\draw[arr] (ligament.south)-- ($(ligament.south) + (-0.2cm,-1.2cm)$) {};

\node[lbl, anchor=south east](lobe_left) at ($(intraop_full.south east) + (-1cm,0.5cm)$)  {Left Lobe};
\draw[arr] (lobe_left.north)-- ($(lobe_left.north) + (0cm,1cm)$) {};

\node[lbl, anchor=south](lobe_right) at ($(intraop_full.south) + (-1cm,0)$)  {Right Lobe};
\draw[arr] (lobe_right.north)-- ($(lobe_right.north) + (0cm,1cm)$) {};

\node[anchor=north west] at (intraop_full.north west) {a)};
\node[anchor=north west] at (intraop_mc.north west) {b)};
\node[anchor=north west] at (intraop_mc_seg.north west) {c)};

\end{tikzpicture}
\caption{Intraoperative surface extraction: a) Merged point cloud from stereo video feed, using poses calculated by OrbSLAM2 and RGB-D data calculated using a disparity matching CNN, b) removing points which were only seen from a single view point, c) segmentation generated by applying a segmentation CNN to each video frame and merging the results into the final point cloud.}
\end{figure}

\begin{figure}
\centering
\begin{tikzpicture}
\node[](r1) at (0,0) {\includegraphics[width=0.75\textwidth]{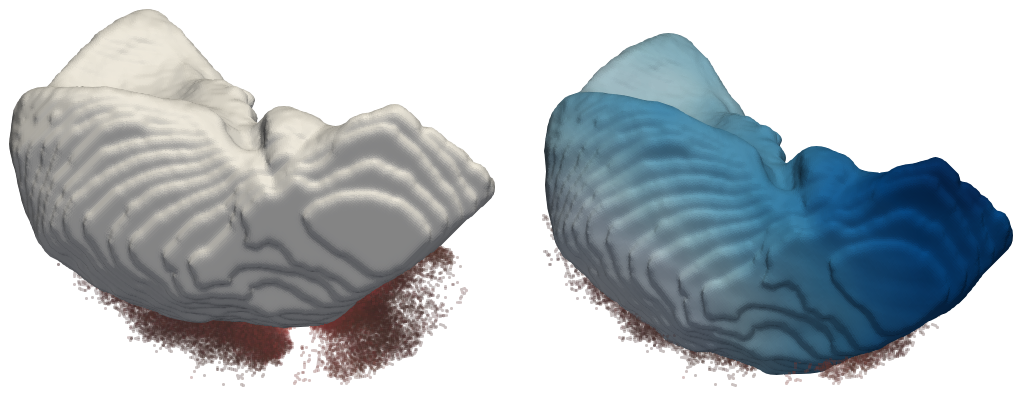}};

\node[anchor=north](r3) at (r1.south) {\includegraphics[width=0.75\textwidth]{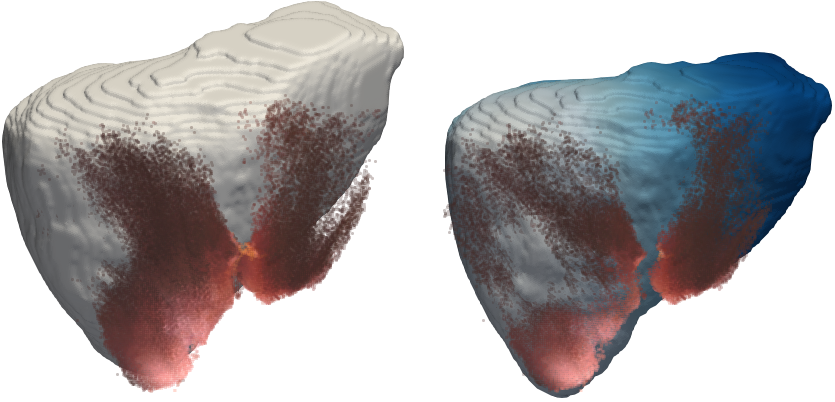}};

\node[anchor=north west] at (r1.north west) {a)};
\node[anchor=north west] at (r3.north west) {b)};

\end{tikzpicture}
\caption{Additional views of the non-rigid registration. The network finds a good overall solution, but has difficulty in the region between the lobes. In this area, the surfaces differ substantially, since there is a gap between the lobes intraoperatively which is not present in the preoperative data (not visible on CT). This causes the network to underestimate the deformation in this region, most likely because it interprets the gap as noise.}
\end{figure}

%
%
%